\pdfoutput=1

\documentclass[11pt]{article}

\usepackage{ACL2023}

\usepackage{times}
\usepackage{latexsym}
\usepackage[size=tiny]{todonotes}
\usepackage{booktabs}
\usepackage{amsfonts,amssymb}

\usepackage[T1]{fontenc}

\usepackage[utf8]{inputenc}

\usepackage{microtype}

\usepackage{inconsolata}

\usepackage{multirow}
\usepackage{hhline}
\usepackage{hyperref}

\usepackage{amsmath}

\usepackage{graphicx}
\graphicspath{ {./image/} }

\usepackage[inkscapeformat=png]{svg}

\usepackage[toc,title,page]{appendix}

\title{
Is there really a Citation Age Bias in NLP?}

\author{Hoa Nguyen \\
  NTT DATA Deutschland SE\\
  \texttt{nguyenviethoa95@gmail.com} \\\And
  Steffen Eger \\
  Natural Language Learning Group (NLLG)\\
  \url{https://nl2g.github.io}\\
  University of Mannheim\\
  \texttt{steffen.eger@uni-mannheim.de}
  \\}

\begin{document}
\maketitle
\begin{abstract}
Citations are a key ingredient of scientific research to relate a paper to others published in the community. Recently, 
it has been noted that there is a citation age bias in the Natural Language Processing (NLP) community, one of the currently fastest growing AI subfields, in that the mean age of the bibliography of NLP papers has become ever younger in the last few years, leading to `citation amnesia' in which older knowledge is increasingly forgotten. In this work, we put such claims into perspective by analyzing the bibliography of $\sim$300k papers across 15 different scientific fields submitted to the popular preprint server Arxiv in the time period from 2013 to 2022.  We find that all AI subfields (in particular: cs.AI, cs.CL, cs.CV, cs.LG) have similar trends of citation amnesia, in which the age of the bibliography has roughly halved in the last 10 years (from above 12 in 2013 to below 7 in 2022), on average. Rather than diagnosing this as a citation age bias in the NLP community, we believe this pattern is an artefact of the dynamics of these research fields, in which new knowledge is produced in ever shorter time intervals.  
\end{abstract}

\section{Introduction}
Biases in citations of scientific papers are 
ubiquitous. 
For example, researchers may disproportionately cite (1) papers that support their own claims \citep{gotzsche2022citation}, (2) papers that have authors with the same gender \citep{lerman:2022}, (3) their own papers \citep{seeber2019self}, or (4) papers of close peers \citep{fister2016toward}. Recently, another citation bias has 
come under investigation, 
namely, `citation amnesia', according to which authors tend to be biased in terms of newer paper, 
`forgetting' the older knowledge accumulated in a scientific field \citep{singh-etal-2023-forgotten,bollmann-elliott-2020-forgetting}. 
Citation amnesia 
has 
been 
discussed especially 
for 
the field of natural language processing (NLP), one of the currently most dynamics subfields of artificial intelligence (AI) \citep{Eger2023NLLGQA,Zhang2023NLLGQA}. For example, \citet{singh-etal-2023-forgotten} find that 
more than 60\% of all citations in NLP papers are from the 5 years preceding a publication and the trend has become considerably worse since 2014; allegedly, current NLP papers are at an ``all-time low'' of citation age diversity. 

In this paper, we take a broader perspective, and examine the age of citations, over time, across different (quantitative) scientific fields. In particular, we examine how the age of the bibliography has developed in the last ten years (from 2013 to 2022) in the science subfields of computer science, physics, mathematics, economics, electrical engineering, quantitative finance, quantitative biology, and statistics. To do so, we leverage arXiv, an extremely popular pre-print server for science, which offers a comparative collection of volumes of papers. We aggregate our different subfields into three classes: (i) AI related papers as a subset of computer science (CS), (2) non-AI CS papers and (3) non-CS papers. We find distinctive trends for the three classes. Non-CS papers have an increasing trend (on average) of citation age in their bibliography: this is expected 
if we assume that papers reference other papers to a large degree 
uniformly across time (in which case the average age of citations will increase as science progresses, as there are older papers to cite each year). CS non-AI papers have a flat trend, i.e., the age of the bibliography has stayed constant across the 10 year period. In contrast, CS AI papers have a strongly decreasing trend, i.e., the age of citations drastically reduces over the ten year period and 
roughly satisfies an exponential decay: e.g., the average age of citations reduces from 12 years in 2013 to below 7 in 2022. This holds true for all four AI subfields we examine: NLP, Computer Vision (CV), Machine Learning and AI proper. Our findings question the previous assessment of `citation amnesia in NLP': instead, it suggests that the (most) dynamic subfields of AI are particularly susceptible to citation age decay and this may especially be a function of the dynamicity of the field. This makes sense: if a field is very dynamic, new knowledge becomes available quickly, 
and past knowledge becomes outdated quickly and cited less frequently. Thus, we believe that the citation amnesia property is a trait exhibited by all very dynamic scientific fields and the fact that citation age patterns have changed in NLP is a property of the changing state of the NLP community \citep{jurgens2018measuring,DidAIGetMoreNegativeRecently,schopf-etal-2023-exploring}.

\section{Related work}\label{sec:related}
Scientometrics studies quantitative characteristics of science. Citations are one of its core concerns. 
%

For instance, \citet{rungta-etal-2022-geographic} show 
that there is a lack of geographic diversity in NLP papers. 
Similarly, \citet{Zhang2023NLLGQA} find a dominance of US industry in most heavily cited AI arXiv papers and an underrepresentation of Europe.  \citet{wahle-etal-2023-cite} show that NLP papers recently tend to disproportionately cite papers within the community itself  

\citet{mohammad-2020-gender} 
study gender gap in NLP research. Other popular aspects of citations investigated in previous work are citation polarity (e.g., is a paper positively or negatively cited) \citep{abu-jbara-etal-2013-purpose,doi:10.1073/pnas.1502280112} and citation intent classification \citep{cohan-etal-2019-structural}. Besides classification, generation for science has recently become popular, including review generation \citep{10.1613/jair.1.12862}, automatic title generation \citep{Mishra2021AutomaticTG,Chen2022TransformersGF} and generation of high-quality scientific vector graphics \citep{Belouadi2023AutomaTikZTS}. 

The papers most closely related to ours are \citet{bollmann-elliott-2020-forgetting} and \citet{singh-etal-2023-forgotten}. \citet{bollmann-elliott-2020-forgetting} look at a 10 year period (2010-2019) and find that more recent papers, published between 2017 and 2019 have a younger bibliography, compared to papers published earlier in the decade. \citet{singh-etal-2023-forgotten} confirm this trend, looking at a larger time frame of publications, encompassing 70k+ papers, showing that NLP papers had an increasingly aging bibliography in the period from 1990 to 2014, but the trend reversed then,\footnote{Somewhat unsurprisingly, 2014 is intuitively the time that the deep learning revolution has gained traction in NLP following papers such as word2vec \citep{NIPS2013_9aa42b31}.} and provide additional analyses. 
In contrast, \citet{DBLP:journals/corr/VerstakASHILS14} show  
with the digital age, older papers also allow to be found more easily, increasing the chance that they will be cited. 
\citet{Parolo_2015} point 
out that the impact of a paper follows a pattern, which increases a year after it is published, reaches its peaks and decreases exponentially. \citet{doi:10.1126/sciadv.1601315} study an interesting relation of a paper's bibliography to its future success: apparently successful papers have low mean but high variance in their bibliography's age distribution.

Our own work connects to the above named as follows: our critical insight is that the age distribution of a bibliography may depend on (1) time and (2) the scientific field considered. Only by setting NLP in relation to other fields can we analyze extents of biases in citation distributions. To do so, we analyze the age distribution of $\sim$300k papers submitted to Arxiv in the last 10 years (2013-2022), spread out across 15 different scientific fields. Looking at arxiv is justified because arxiv has become an extremely popular preprint server for science since its dawn in the early 1990s\footnote{See \url{https://info.arxiv.org/help/stats/2021_by_area/index.html}.} that hosts several of the most influential science papers \citep{Eger2023NLLGQA,Zhang2023NLLGQA,Clement2019OnTU,Eger2018PredictingRT} made available at much faster turnaround times than in traditional conferences or journals. 



\section{Dataset}
\label{sec:dataset}
We describe the source from which we extract our dataset\footnote{\url{https://drive.google.com/drive/u/0/folders/1k0GOvi9-m5Hrs4EO6O6iuskJOV7Oq5cl}} and the steps we perform to construct our dataset, which we make available at \url{https://github.com/nguyenviethoa95/citationAge}. 

\paragraph{Data Source} We create our dataset leveraging \textit{arXiv} and Semantic Scholar. 
\textit{\textit{arXiv}}\footnote{\url{https://arXiv.org/}} is an extremely popular open access pre-print server focusing on `hard sciences' 
like mathematics, physics and computer science, 
along with other quantitative disciplines such as biology and economics.  
It currently hosts more than two million articles in eight subject areas. 
\textit{Semantic Scholar}\footnote{\url{https://www.semanticscholar.org/}} is a free and open access database developed by the Allen Institute for Artificial Intelligence. 
It employs machine learning technology to index scientific literature, extract the metadata from the paper content, and perform further analysis on the metadata. As of January 2023, the number of records in Semantic Scholar is more than 200 million, which includes 40 million papers from computer science disciplines.

\paragraph{Subcategory Selection}
For computational reasons, we do not focus on the whole of \textit{arXiv} but only on manageable subsets. 
\textit{arXiv} papers are sorted into eight main categories: computer science, economics, electrical engineering, math, physics, quantitative biology, quantitative finance and statistics.\footnote{See \url{https://arXiv.org/category_taxonomy}.}
Each category is further divided into sub-categories, e.g., cs.CL stands for computation \& language (NLP) within the computer science main category. 
For each of the main categories, we choose the subcategories containing the highest number of papers, see the appendix. 
An exception is the main category of computer science, which is our focus.
In particular, along with cs.CL,  we also choose seven other sub-categories from CS.  
We distinguish (1) \textbf{cs-non-ai} from (2) \textbf{cs-ai} papers. 
The latter contain papers submitted to AI related fields (Computer Vision, AI, NLP, Machine Learning), the former contains papers submitted to non-AI related fields (such as data structures and algorithms). 

\begin{table*}
\footnotesize
\centering
\begin{tabular}{c*{5}{c}}
\toprule
Category                   & Subcat. &  Description & Full-name                                & Number   & Total                   \\ \midrule
\multirow{4}{*}{cs-non-ai} & cs.CR    & Computer Science   & Cryptography and Security               & $14,741_{14,531}$  & \multirow{3}{*}{$60,346_{59,822}$}    \\ 
                           & cs.IT    & Computer Science     & Information Theory                    & $23,965_{23,845}$  &                         \\ 
                           & cs.NI    & Computer Science    & Networking and Internet Architecture                 & $10,888_{10,786}$ &                         \\ 
                           & cs.DS    & Computer Science    & Data Structures and Algorithms          & $10,752_{10,660}$   &                         \\ \midrule
\multirow{4}{*}{cs-ai}     & cs.AI     & Computer Science    & Artificial Intelligence                 & $13,529_{8316}$ &   \multirow{4}{*}{$139,769_{110,024}$}                       \\ 
                           & cs.CV     & Computer Science    & Computer Vision and Pattern Recognition & $65,685_{48,391}$ \\ 
                           & cs.LG      & Computer Science   & Machine Learning                        & $57,935_{29,688}$ &                         \\ 
                           & cs.CL      & Computer Science   & Computation and Language                & $30,867_{23,629}$ &                         \\ \midrule
\multirow{4}{*}{non-cs}    & math.AP  & Mathematics   & Analysis of PDEs                            & $32,530_{32,229}$    \\ 
                           & econ.GN   & Economics  & General Economics Economics                     & $2112_{885}$     &                         \\ 
                           & eess.SP      & Electrical Engineering     & Signal Processing                & $12,505_{12,435}$    & \multirow{3}{*}{$108,288_{102,532}$}                 \\ 
                           & hep-ph  & Physics & High Energy Physics - Phenomenology  & $47,364_{43,331}$     &                         \\ 
                           & q-bio.PE  & Quantitative Biology    & Populations and Evolution                    & $4797_{4708}$     &                         \\ 
                           & q-fin.ST  &  Quantitative Finance      & Statistical Finance                & $1238_{1231}$     &                         \\ 
                           & stat.ME   & Statistics   & Methodology                      & $13,775_{13,667}$  &                         \\ \bottomrule
\end{tabular}
\caption{\label{statistics-table}Dataset statistics of sub-categories in our dataset. The numbers in subscripts are the actual numbers of publications in our dataset (timeouts in querying SemanticScholar may result in lower actual numbers).}
\label{tab:stastistics}
\end{table*}

\paragraph{Data Collection}
We collect papers within the period of 10 years between January 2013 and December 2022. 
Thereby, we make use of the \textit{arXiv} dataset hosted by kaggle\footnote{\url{https://www.kaggle.com/datasets/Cornell-University/arXiv}}, 
which offers an easier way to access metadata of the actual corpus. The metadata consists of relevant attributes of a scholarly paper such as title, authors, categories, abstract, and date of publication. However, the reference papers from the bibliography are not listed in this metadata. 

Thus, we extract the list of references from Semantic Scholar. 
In particular, we use the \textit{arXiv} ID to query the Semantic Scholar  API\footnote{\url{https://www.semanticscholar.org/product/api}}, search for the paper and retrieve the list of reference papers in the bibliography. Importantly, each paper can be assigned to multiple categories, however, we only use the \textbf{primary category} to sort papers into our dataset. 

\begin{figure}[!htb]
\centering
\includegraphics
[scale=0.60]{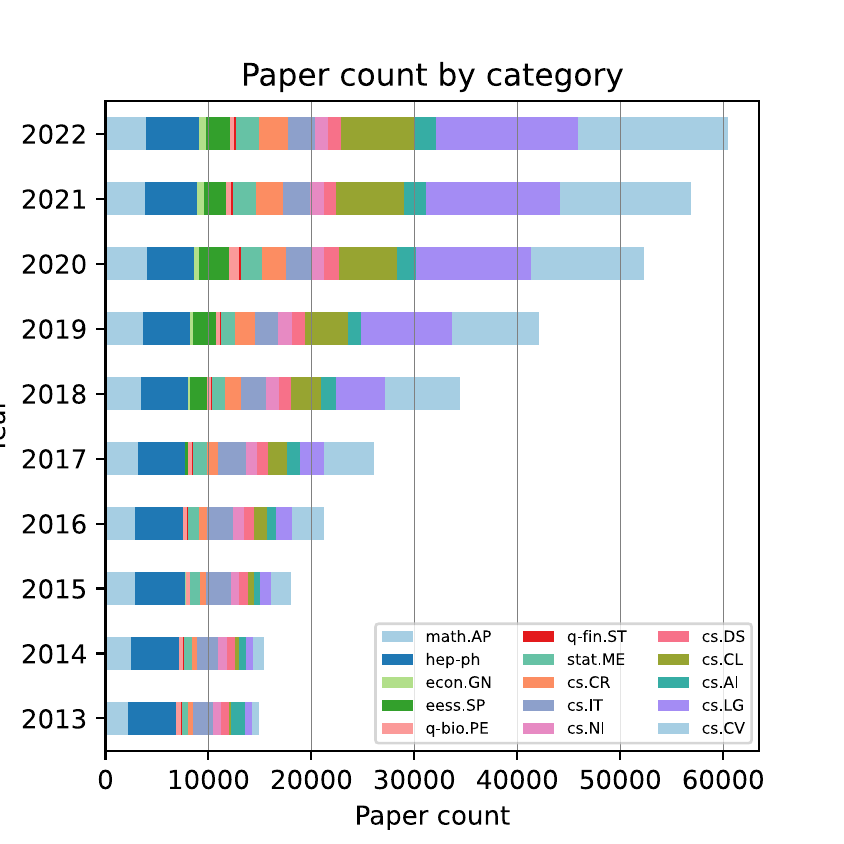}
\caption{Number of papers published from 2013 to 2022 by category. See Table \ref{table:cs_submissions} and Table \ref{table:noncs_submissions} for the detailed submission of each subcategory.}
\label{fig:all_num_year}
\centering
\end{figure}

\begin{figure}[!htb]
\centering
\includegraphics[scale=0.60]{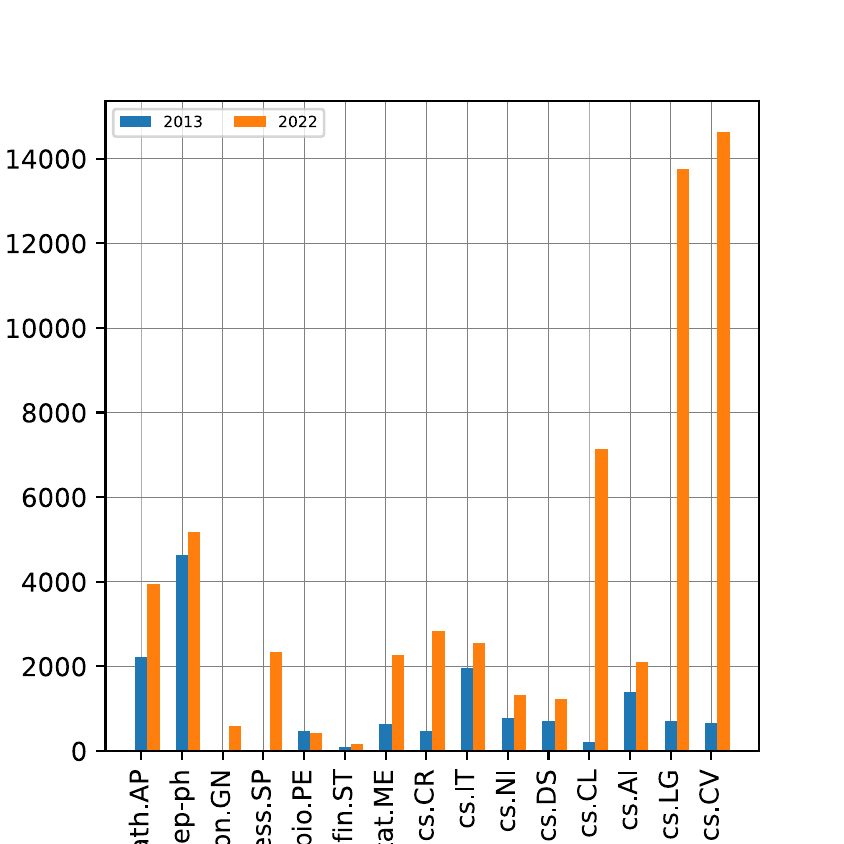}
\caption{Paper count by category in 2013 and 2022. See Tables  \ref{table:cs_submissions} and 
\ref{table:noncs_submissions} for 
exact numbers.}
\label{fig:all_num_by_year_1322}
\centering
\end{figure}

\paragraph{Data statistics}
Our final dataset comprises 
8 main categories with 15 sub-categories of scientific papers along with their metadata and their corresponding list of references in the period from January 2013 until December 2022. Our dataset is summarized in Table \ref{tab:stastistics}. We notice that computer science, mathematics, and physics 
attract the largest number of paper submissions by far.  Also, the total number of CS AI submissions (139k) 
is more than double the non-AI related CS submissions (60k), as shown in Table \ref{tab:stastistics}. In 2022, the number of CS AI papers (37,626) is considerably more than the number of non-AI CS papers (6,752) and non CS papers (19,297) combined, see Figure \ref{fig:all_num_year} and Tables \ref{table:cs_submissions} and \ref{table:noncs_submissions}. 
The same is not true for earlier time periods, e.g., in 2013, there were only $\approx3k$ CS AI submissions but $\approx4k$ CS non-AI submissions and $\approx8k$ non-CS submissions. This indicates that AI has been growing most strongly in our data. Figure \ref{fig:all_num_by_year_1322} demonstrates the difference in the development of research output in our dataset by plotting the numbers of papers submitted to arXiv in 2013 and 2022. We observe that the most fast growing fields are indeed computer science fields with AI focus. 
Among the AI related field, cs.CL (Computer Linguistics) has the highest growth rate of almost $\approx$ 32-times (219 submissions in 2013 to above 7k submissions in 2023), followed by cs.CV (Computer Vision) at $\approx$ 22-times and cs.LG (Machine Learning) at $\approx$ 20-times, see Figure \ref{fig:all_num_by_year_1322} and Tables \ref{table:cs_submissions} and \ref{table:noncs_submissions}. We note that econ.GN and eess.SP have very low support for the years 2013 to 2017, making statistics on them more unreliable.

\section{Analysis}
\label{sec:analysis}

\begin{table*}[htb]
\centering
\begin{tabular}{c|cccc|cccc}
     \toprule
     \textbf{year} & \textbf{cs.CR} & \textbf{cs.IT} & \textbf{cs.NI} & \textbf{cs.DS}  & \textbf{cs.AI} & \textbf{cs.CL} & \textbf{cs.CV} & \textbf{cs.LG} \\
     \midrule 
     2013    &  9.71	& 9.7 &	7.33 &	12.95 &	17.61 &	10.9 &	9.54 &	10.91  \\ 
     2014  &  9.37	& 9.83	& 7.27	& 13.15	& 12.21	& 9.94	& 9.09	& 9.95 \\ 
          2015  &   8.85	& 9.85	& 6.87	& 13.36	& 10.82	& 8.52	& 7.73	& 9.32  \\ 
          2016  &   9.11	& 9.81	& 7.14	& 13.25	& 9.91	& 7.68	& 8.67	& 8.76\\ 
          2017  &   8.31	& 9.77	& 6.97	& 13.37	& 9.36	& 7.43	& 6.8	& 8.44\\ 
          2018  &  7.88	& 9.18	& 6.74	& 13.52	& 8.72	& 6.78	& 6.1	& 7.96  \\ 
          2019  &  7.88	& 9.82	& 6.76	& 14.33	& 8.74	& 6.59	& 6.02	& 7.83   \\ 
          2020  &   7.39	& 9.5	& 6.85	& 14.37	& 8.31	& 6.3	& 5.94	& 7.68  \\ 
          2021  &  7.47	& 9.76	& 6.66	& 14.08	& 7.33	& 6.13	& 5.82	& 7.46 \\ 
          2022  &  7.59  &  10.01	& 6.77	& 14.23	& 7.24	& 6.15	& 5.93	& 7.61 \\ 
     \bottomrule 
\end{tabular} 
\caption{Left: Mean AoC cs-non-ai categories. Right: cs-ai categories.}
\label{table:csmean}
\end{table*}

\begin{table*}[htb]
\centering
\begin{tabular}{c|ccccccc}
     \toprule
     \textbf{year} & \textbf{q-bio.PE} & \textbf{q-fin.ST} & \textbf{stat.ME} & \textbf{hep-ph}  & \textbf{math.AP} & \textbf{econ.GN} & \textbf{eess.SP} \\
     \midrule 
     2013    & 13.98  &  	13.29  &  	13.26 &  10.34  &  14.93  &  0  &  9.03\\ 
     2014  &   13.28  &  13.36  &  13.43  &  10.96  &  15.21  &  13.2  &  16.12\\ 
          2015  & 14.6  &  13  &  13.64  &  11.19  &  15.09  &  14.47  &  5.53 \\ 
          2016  & 15.01  &  14.59  &  13.68  &  11.56  &  15.33  &  23.27  &  10.96 \\ 
          2017  & 14.62  &  15.41  &  14.09  &  12.25  &  15.74  &  10  &  9.59 \\ 
          2018  & 14.79  &  14.09  &  14.54  &  12.05  &  16.05  &  14.98  &  9.28\\ 
          2019  & 15.17  &  14.26  &  14.73  &  12.41  &  16.29  &  13.43  &  8.45\\ 
          2020  & 11.68  &  12.62  &  14.3  &  13.08  &  16.48  &  12.62  &  8.12\\ 
          2021  & 12.81  &  12.59  &  14.4  &  13.14  &  16.5  &  12.59  &  8.33\\ 
          2022  & 14.61  &  11.71  &  14.68  &  13.66  &  17.17  &  11.71  &  8.22\\ 
     \bottomrule
\end{tabular}
\caption{Mean AoC of non-cs categories.}
\label{table:noncsmean}
\end{table*}

\begin{figure*}[!htbp]
\centering
\includegraphics[scale=0.25]{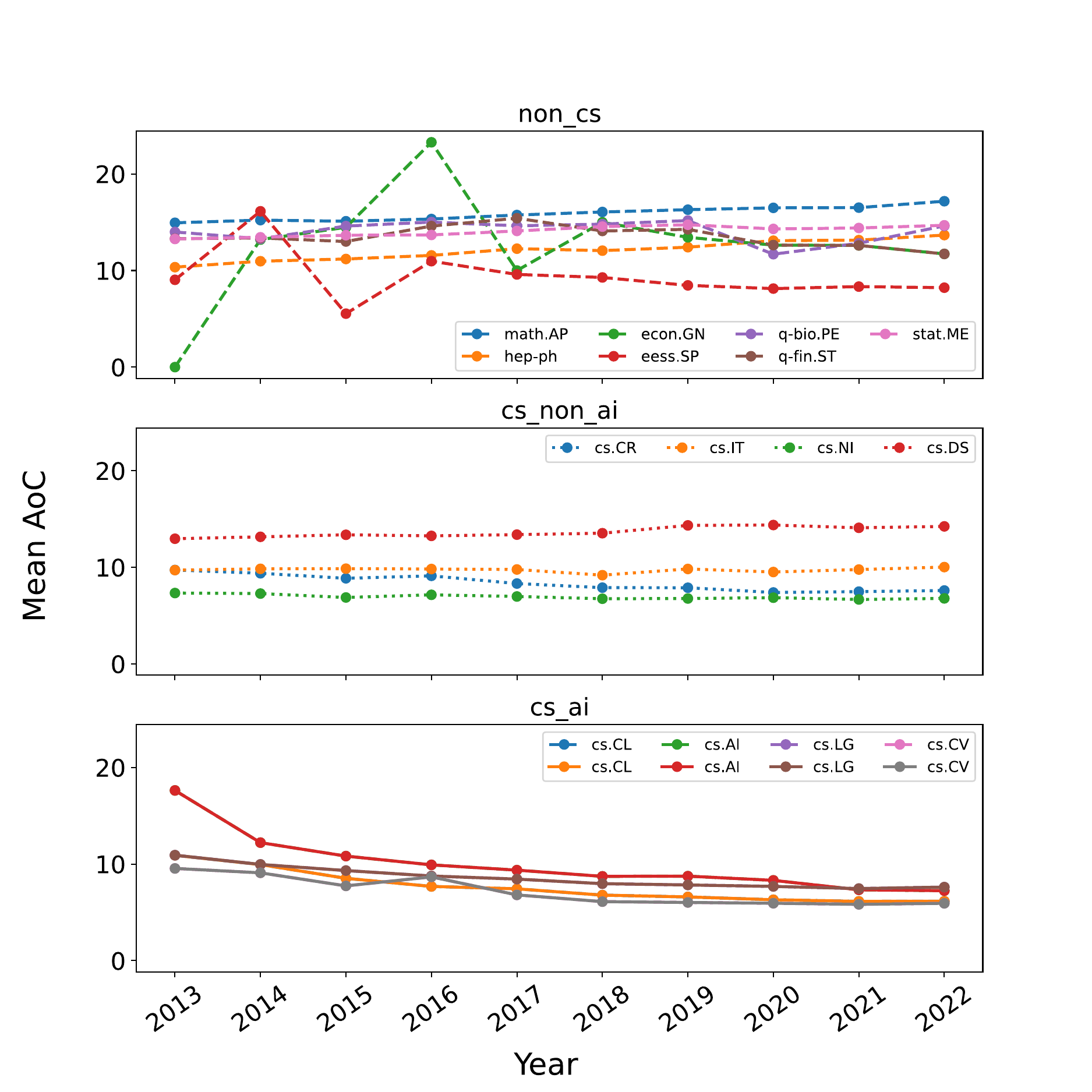}
\includegraphics[scale=0.25]{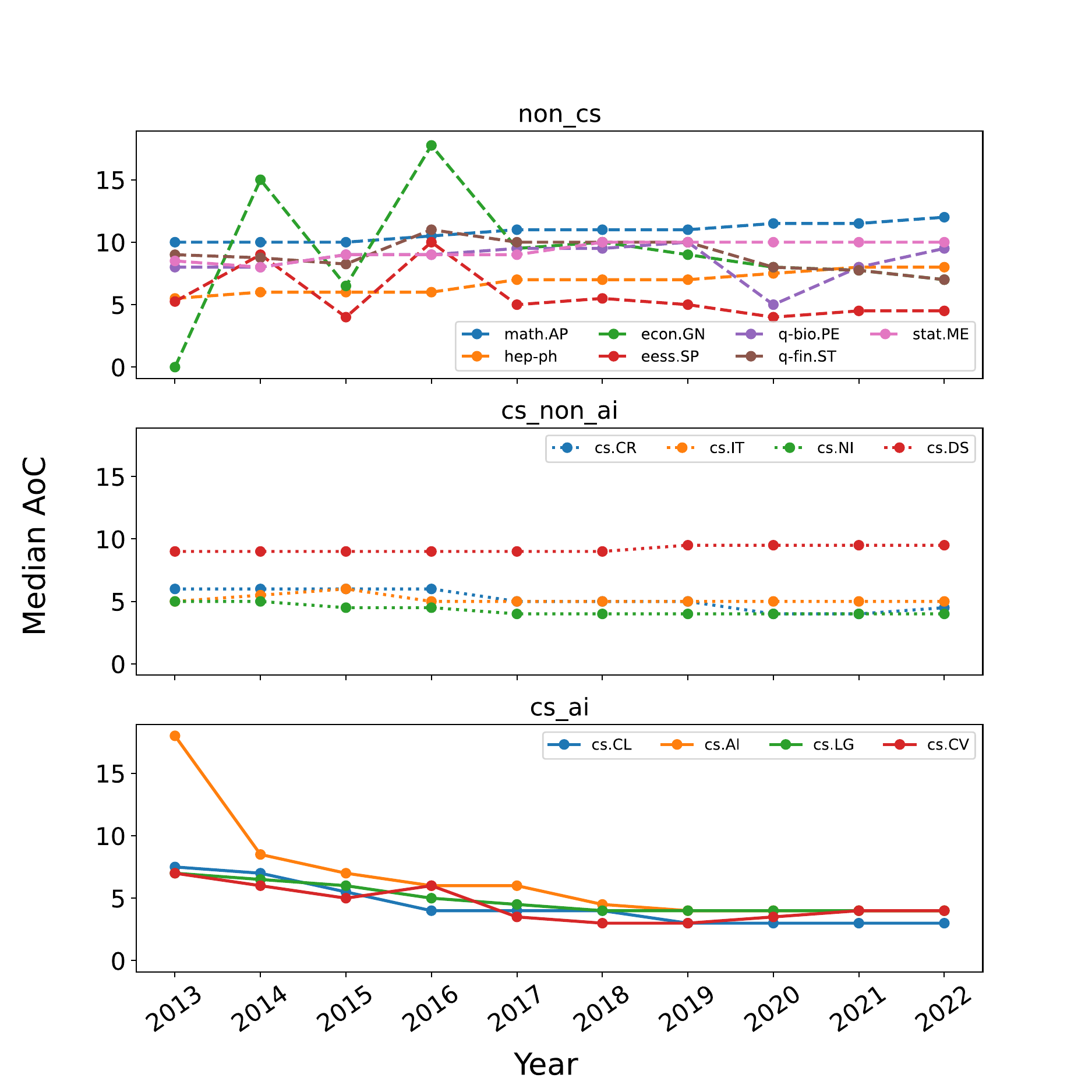}
\caption{Mean (left) and median (right) age of citation by categories. Tables \ref{table:csmean}, \ref{table:noncsmean}, \ref{table:csmedian} and \ref{table:noncsmedian} give 
exact 
numbers.}
\label{fig:mean_aoc}
\centering
\end{figure*}

In this section, we use the dataset 
constructed in Section \ref{sec:dataset} to perform different temporal analyses on the references of scientific papers. In particular, we focus on investing how the gap between a cited paper and the original papers has changed over a decade from 2013 to 2022. 

\subsection{Metrics}
\label{sec:metrics}
To examine the change in the trends of referencing of old papers, we  use the  metrics described below. Our notation is inspired by \citet{singh-etal-2023-forgotten}.

\paragraph{Age of Citation}
The age AoC of 
a citation $y_i$ in a paper $x$ can be defined 
as the difference between the year of publication (YoP) of both: 
\[ \textit{AoC}(x, y_i) = \textit{YoP}(x) - \textit{YoP}(y_i) \]
Using this, we calculate the mean age of the 
$M$ references 
of a paper $x$ as:
\begin{align*}
  \overline{\textit{AoC}}(x) = \frac{1}{M} \sum_{i=1}^{M}  \textit{AoC}(x, y_i)
\end{align*}
Finally, when we have $N$ papers $x_j$ published in a year $t$, we calculate the average over all $N$ papers to obtain the mean citation age in year $t$: 
\begin{align*}
_m AoC(t) = \frac{1}{N}\sum_{j=1}^{N} \overline{\textit{AoC}}(x_j)
\end{align*}
\paragraph{Percentage of old citations} 
We calculate the percentage of old citations
as the percentage of the `old' (published at least $k=10$ years before the citing paper) references
in a paper: 
\begin{align*}
\textit{$PoOC$}(x) = 
\frac{|\mathfrak{O}_k(x)|}{M}
\end{align*}
where $\mathfrak{O}_k(x) = \{y\,|\, 
\textit{AoC}(x,y) 
\ge k\}$ is the set of references whose publication age is $k$ years older than that of the citing paper $x$. From this formula, we can again compute the mean percentage of old papers over any given year $t$ with $N$ papers, as follows:
\begin{align*}
\textit{$_m PoOC$}(t) =  
\frac{1}{N}\sum_{x=1}^{N}
\textit{$PoOC$}(x)
\end{align*}


\subsection{Mean and median age of citations} 
To examine the change in the age of cited papers over the different fields, we calculate the mean age of the papers by year and plot this in Figure \ref{fig:mean_aoc} and Tables \ref{table:csmean} and \ref{table:noncsmean}. 
There is a large discrepancy between the mean age of citations 
across different categories. 

For example, cs.CL (NLP) has decreased from $_m\textit{AoC}= 
10.9$
in 2013 to $_m\textit{AoC}= 6.15$  in 2022 --- a decrease of  44\%. 
The other AI related fields show similar decreases: cs.AI has decreased by 
59\% 
from $_m\textit{AoC}(2013)= 17.61$ in 2013 to $_m\textit{AoC}(2022) = 7.24$ in 2022, cs.CV by 
38\% 
from $_m\textit{AoC}(2013)= 9.54$ to $_m\textit{AoC}(2022) = 5.93$ and cs.LG by 
30\% from $_m\textit{AoC}(2013) = 10.91$ to $_m\textit{AoC}(2022) = 7.61$. 
\textbf{The average decrease of mean age of citations for CS AI categories between 2013 and 2022 is 43\%. The average yearly rate of decrease in CS AI categories is 6\%;}\footnote{By this, we mean the average over the ratios $\frac{y_t}{y_{t-1}}-1$ where $y_t {=} {_m\textit{AoC}(t)}$, for $t=2014,\ldots,2022$.}
in other words, the age of citations in a typical CS AI paper decreases by 6\% on average from year to year, in the indicated time frame. 
In contrast, the four non-AI CS fields in our collection have a maximum decrease of 
22\% (cs.CR) and two out of four fields have even a small increase (cs.IT and cs.NI) of up to 
10\%.
\textbf{The average decrease of mean age of citations between 2013 and 2022 for CS non-AI categories is 4\%; the average yearly rate of decrease is 0.5\%}.
Concerning the non-CS fields, 
4 out of 7 
show an increase in citation age between 2013 and 2022 (q-bio.PE, stat.ME, math.AP, 
hep-ph). 
\textbf{The average decrease of mean age of citations for non-CS categories between 2013 and 2022 is -4\%} (i.e., an increase of 4\%) \textbf{and the average yearly rate of decrease is -3\%}.
Similarly, Figure \ref{fig:mean_aoc} depicts the median age of citations --- the median is less affected by outliers. We observe the same pattern as for the mean, indicating that outliers do not influence our results. In fact, the Pearson correlation between CS categories is 93\% (median vs.\ mean) and it is 88\% for non CS categories. The decreases in CS AI categories are more extreme: on average, the yearly rate of decrease in AoC is 8\%, while it is 1\% for CS non-AI categories. For non-CS categories, it is -4\%.

\begin{figure}[htb]
\centering
\includegraphics[scale=0.50]{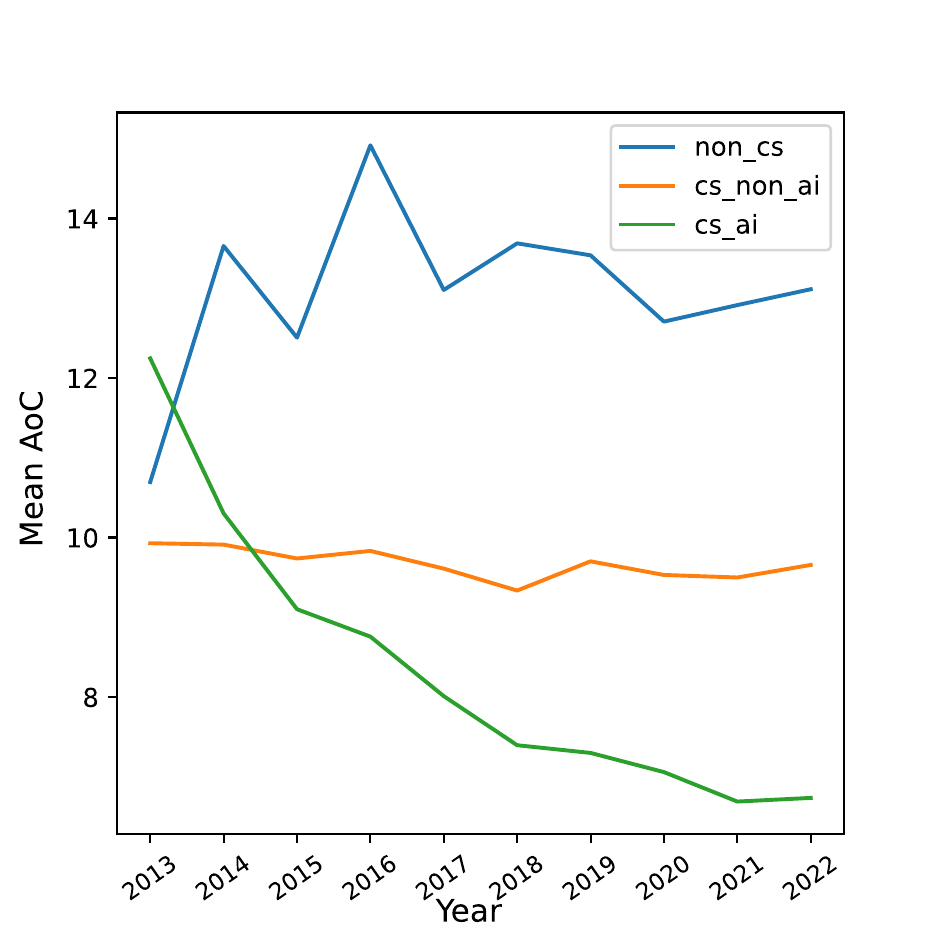}
\caption{Mean AoC of papers published from 2013 to 2022 grouped by general groups: CS AI, CS non-AI and non-CS.}
\label{fig:all_mean}
\end{figure}

Figure \ref{fig:all_mean} shows the bibliography age dynamics from 2013 to 2022 averaged over CS AI, CS non-AI and non-CS papers.

\subsection{Percentage of old citations} 
\begin{figure}[!htb]
\centering
\includegraphics[scale=0.25]{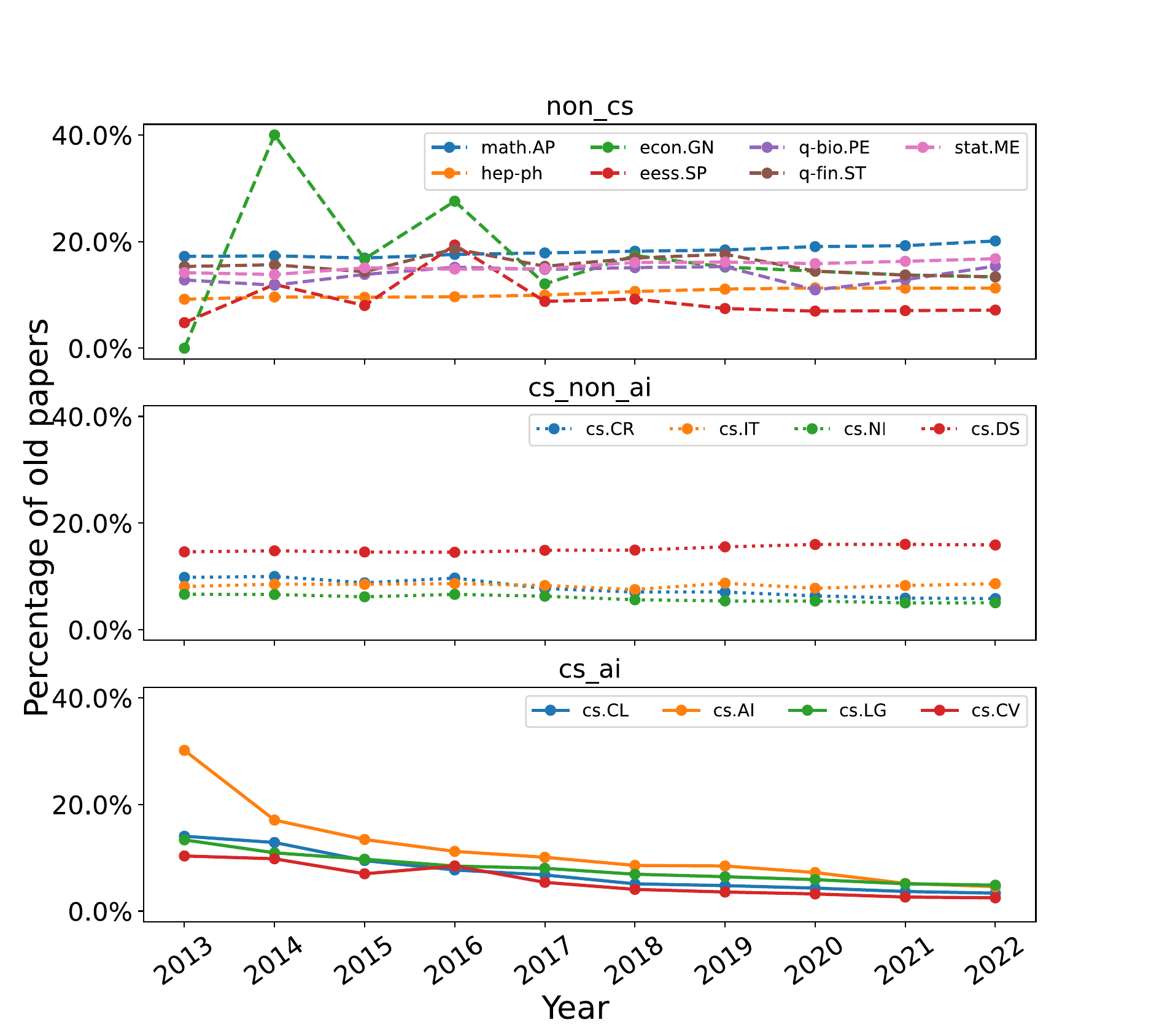}
\caption{Percentage of old paper by categories and year. See Table \ref{table:csold} and Table \ref{table:noncsold} for details.}
\label{fig:percentage_old_paper}
\centering
\end{figure}

The percentage of old papers follows a similar trend across our three high-level categories: cs-ai fields have decreased by 
75\% 
on average between 2013 and 2022 in terms of the proportion of old citations; cs-non-ai fields have decreased by 
33\% 
and non-cs fields have 
decreased by -7\%. The Pearson correlation between CS categories is 88\% (mean AOC vs.\ PoOC) and that of non-CS categories is 72\%.  
For example, cs.CL had 14\% of all citations as old citations in 2013, but below 4\% in 2022. cs.AI is again the most extreme: it decreases from 30\% in 2013 to below 5\% in 2022. Details can be found in Tables \ref{table:csold} and \ref{table:noncsold}. 

\subsection{Mean citation age of influential papers}

\begin{figure}[!htb]
\centering
\includegraphics[scale=0.25]{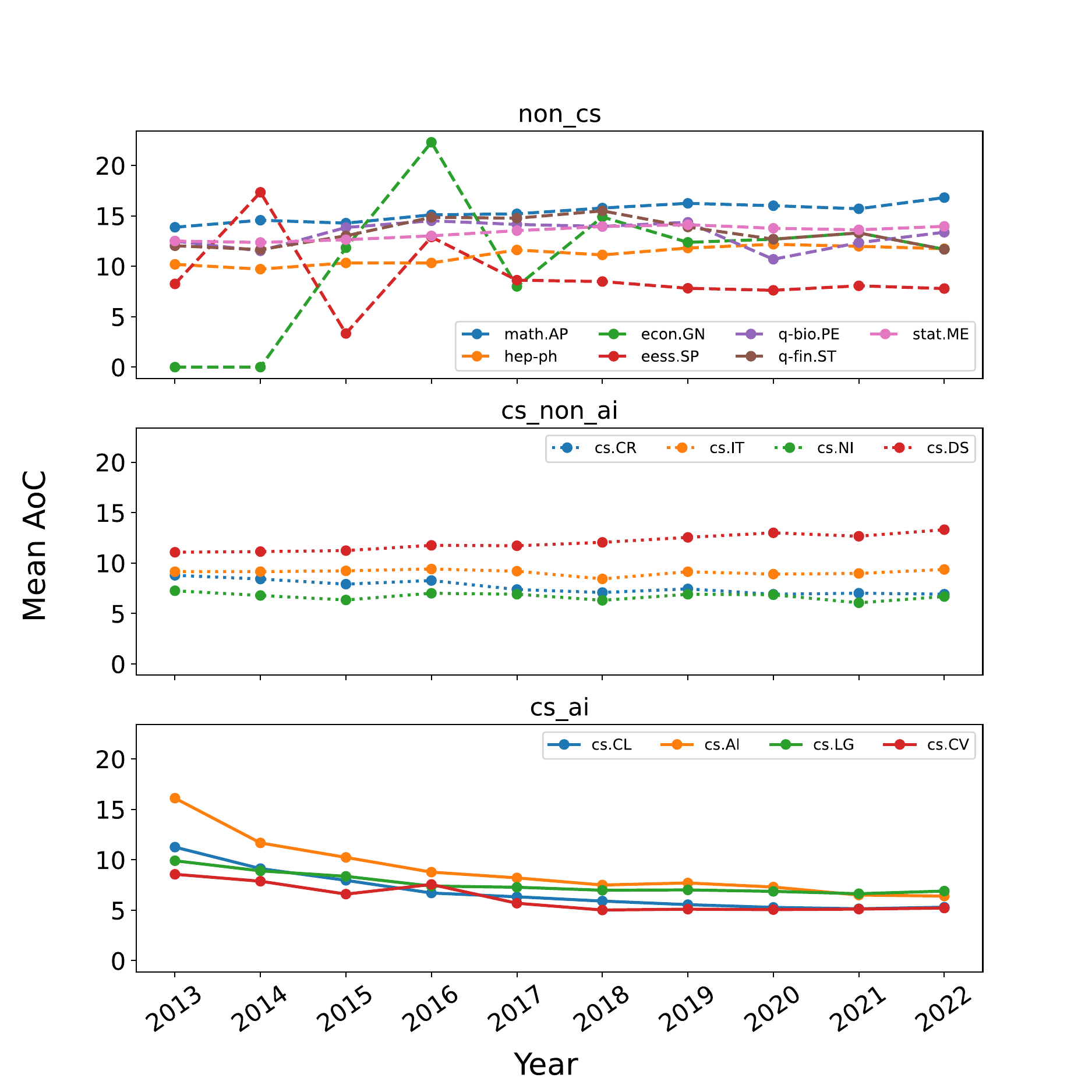}
\caption{Mean AoC of influential paper by categories and year. See Table \ref{table:csinfluential} and Table \ref{table:noncsinfluential} for details.}
\label{fig:mean_of_old_infl}
\centering
\end{figure}

In addition, we investigate how the age of the \emph{influential} references cited in a paper has changed over our time period.
A citation is considered ``highly influential'' if it has major impact on the
citing paper.  The identification of these ``highly influential'' papers is done based on machine learning algorithms developed by \textit{Semantic Scholar}, which uses multiple criteria for calculation.  The major criterion is the number of times the citation occurs in the full text and the surrounding text around the citation.  Here, we calculate the mean of old citations within the ``highly influential'' citations. Figure \ref{fig:mean_of_old_infl} plots the temporal change of the age difference between the influential citations within a publication and the publication itself.
 
Firstly, the mean AoC of influential citations 
is typically lower than the normal mean AoC in all fields and subcategories over the years. 
For example, 
cs.CV has $_m\textit{AoC}(2013)= 9.54$ and $_m\textit{AoC}(2022)= 5.93$, while its influential mean AoC  are $_m\textit{AoC}(2013)= 8.56$ and $_m\textit{AoC}(2022)= 5.10$, which are lower than the normal mean AoC of the same year. 
On average, the influential citations are 0.8 years younger than the average citations. 
This makes intuitively sense: the references that really influence a given paper are more recent. 
Secondly, the temporal changes of the mean AoC of influential citations of all fields is similar to the changes of mean AoC of all citations. 
For example, the mean age of citations in CS AI categories has decreased by 
46\% 
on average between 2013 and 2022 (it is worth pointing out that the decrease has slowed down, however, in recent years), 
the CS non-AI categories have largely remained unchanged (decrease of 2\%), and the non-CS categories have 
decreased by -6.5\%.  
Details can be found in Tables \ref{table:csinfluential} and \ref{table:noncsinfluential} in the appendix.

\section{Discussion}\label{sec:discussion}
Our results --- regarding the mean and median age of (influential) citations as well as the percentage of old citations --- in the previous section all point in the same direction: the age of the bibligraphy in the CS AI subfields we examined has considerably decreased 
over the years we considered. 
In this respect, the AI subfields behave 
very 
differently from non-CS and non-AI fields. 

\begin{figure}[!htb]
    \centering
    \includegraphics[scale=0.35]{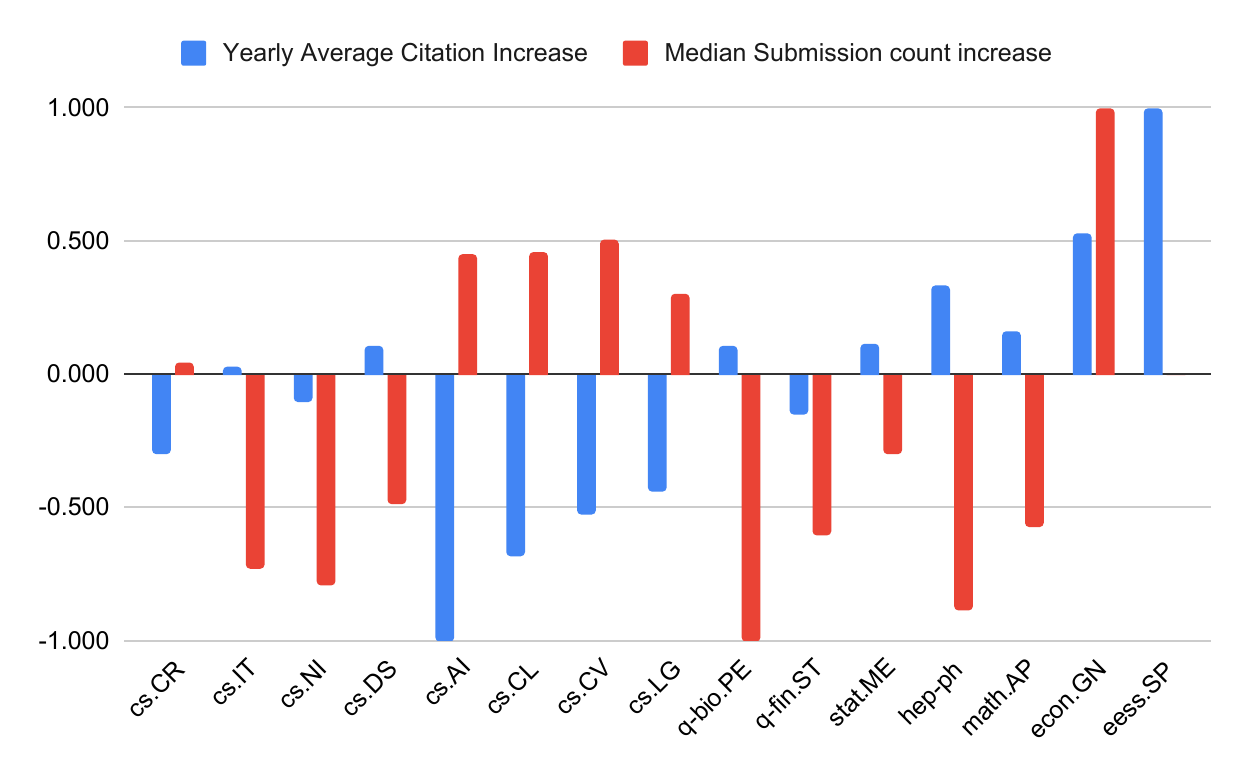}
    \caption{Yearly average age of citation increase vs.\ yearly submission count increases per field, normalized to $[-1,1]$.}
    \label{fig:dynamics}
\end{figure}

We illustrate the differences between the fields we consider in Figure \ref{fig:dynamics}. 
There, 
we plot the yearly average citation  increases (negative numbers denote decreases) vs.\ the median yearly submission increases of each field; the latter is an 
indicator of the dynamicity of the field. 
We normalize all numbers to $[-1,+1]$ to ensure comparability using $x\mapsto 2\frac{x-\min{\mathbf{x}}}{\max{(\mathbf{x}})-\min{(\mathbf{x}})}-1$. CS AI fields have clearly distinct patterns: they have high decreases in yearly average age of citations and high yearly increases of submission numbers to arxiv. 
The more established CS fields are less dynamic: their submission numbers grow slowly or even decrease over the decade considered and, simultaneously, their 
bibliography age is also relatively stable over the time period. 
Non-CS fields typically have positive yearly average age of citation increases and most strongly decreasing submission numbers (e.g., hep-ph and math.AP have largely stagnated in the last few years or slightly decreased); an exception are eess.SP and econ.GN. We note, however, that (1) these two subfields have comparatively low numbers of submissions, making statistics  less reliable, and (2) it may not have been common (e.g.) in economics to submit papers to arxiv before 2018, so increases in submissions may actually not reflect the dynamicity of the field but 
behavioral changes in that community.

From this broader perspective, it is unclear whether there is a citation age bias \emph{specifically} in NLP. 
Our results indicate that NLP is simply another field like other AI fields 
which are all characterized by high dynamicity, i.e., many newly incoming researchers (and submissions) and quickly changing state-of-the-art solutions.\footnote{A case in point is the area of evaluation metrics in NLP, which has been dominated by models developed in the early 2000s \citep{papineni-etal-2002-bleu,lin-2004-rouge} for a long time, but has then been quickly superseded by a much higher-quality class of metrics since the late 2010s \citep{zhao-etal-2019-moverscore,Zhang2020BERTScore:,rei-etal-2020-comet,sellam-etal-2020-bleurt,chen-eger-2023-menli} whose high citation rates document the community's fast \& wide-scale adoption in recent years.} 
In such an environment, the observed changes in the age of the bibliography may simply be a `natural' response. 
\section{Concluding remarks}
We examined the age of the bibliography across 15 different scientific fields in a dataset of papers submitted to Arxiv in the time period from 2013 to 2022. We found that the dynamic AI fields are all affected by a decreasing age of bibliography over the considered time period, while more established fields do not show the same trend. We believe that this trend is very natural: for example, according to \url{https://aclweb.org/aclwiki/Conference_acceptance_rates}, the submission rates to the main ACL conference(s) have increased five-fold between 2013 and 2022, from 664 submitted papers to 3378 papers. Thus, from the viewpoint of 2013 the year 2022 can be perceived of encompassing ``5 years''. If we take this increase in submissions and money invested into account,\footnote{See \href{here}{
https://www.goldmansachs.com/intelligence/pages/ai-investment-forecast-to-approach-200-billion-globally-by-2025.html}}, especially from the big US AI companies \citep{Zhang2023NLLGQA}, it is clear that the age of citations must become younger. While we expect that 2023 has seen 
additional rejuventation of the bibliography, mainly due to ChatGPT and the LLM revolution \citep{Bubeck2023SparksOA,Leiter2023ChatGPTAM}, our numbers and graphs appear to imply that this trend of decreasing age of citations may soon reach a bottom: for example, there is only a marginal difference in the mean age of citations in the four AI fields we considered between 2020, 2021, and 2022 --- such a pattern is expected in exponential decays, in which the rate of decrease is proportional to the current value. 

We thus want to express a word of caution in interpreting statistical trends as bias (that pertain to particular communities), a tendency that may be fueled by the NLP community's increasing 
self-absorbedness and in-group bias \citep{wahle-etal-2023-cite}. 

Future work should look at the age of citations in more scientific disciplines, published in varying outlets, and across larger time frames. Future work should also develop statistical models of the age of citations in a paper's bibliography to determine \emph{statistical bias}, defined as the deviation from the expected value. 

\section*{Limitations}
We (and others) obtain citation information from SemanticScholar, but we observe that this engine --- like other engines --- is error-prone. 
For a quality check, we manually verify a random subset of our dataset and compare the reference list of data from \textit{SemanticScholar} 
to the manually annotated references. 
We identify 
some of the common error made by \textit{SemanticScholar} as follows. (a) Missing reference: the reference 
in the paper is missing from the list 
provided by \textit{SemanticScholar}. (b) Wrongly assigned reference: The reference listed by \textit{SemanticScholar} does not match with the reference listed in the full-text. Moreover, we notice that the errors do not occur equally in all types of publications. 
For instance, publications 
from large international conferences and journals seemingly may not suffer as much. 
Additionally, older publications 
also seem to suffer more heavily. 
This may be due to the \emph{SemanticScholar} parsing algorithm, which may be trained on tuned on particular data. 
Other limitations 
relate to the Kaggle arxiv snapshot which may not contain all arxiv papers. 

We believe, however, that our results are trustworthy, because 
individual errors 
tend to cancel out on an aggregate level, which we  
exclusively report in our work. Furthermore, even 
medium-size subsets of data 
are often sufficient to report accurate aggregate statistics.

\section*{Acknowledgements}
The NLLG group is supported by the BMBF grant Metrics4NLG as well as the DFG Heisenberg grant EG 375/5--1.

\bibliography{anthology,custom}
\bibliographystyle{acl_natbib}

\appendix

\begin{appendices}

\textbf{Mathematics}
{math.KT: 2357, math.HO: 2730, math.GN: 3041, math.GM: 3517, math.CT: 3722, math.SP: 4442, math.SG: 4545, math.MG: 5774, math.OA: 7677, math.AC: 7896, math.AT: 8375, math.QA: 8639, math.LO: 9016, math.CV: 9917, math.RA: 10173, math.GR: 13281, math.ST: 13682, math.RT: 14256, math.CA: 14420, math.GT: 14804, math.FA: 18708, math.DS: 21177, math.NA: 25910, math.DG: 27471, math.OC: 28331, math.NT: 30311, math-ph: 30693, math.AG: 33929, math.PR: 36838, math.CO: 42004, \textbf{math.AP: 45244} }

\textbf{Physics}
{nlin.CG: 492, physics.atm-clus: 1194, physics.pop-ph: 1284, physics.space-ph: 2070, nlin.AO: 2575, physics.ed-ph: 2880, physics.hist-ph: 2982, physics.data-an: 3168, physics.ao-ph: 3229, physics.geo-ph: 3551, nlin.PS: 4156, physics.med-ph: 4190, physics.class-ph: 4580, nlin.SI: 4929, physics.acc-ph: 5883, physics.bio-ph: 5910, nlin.CD: 6334, cond-mat.other: 6667, physics.comp-ph: 7298, physics.app-ph: 8754, physics.gen-ph: 8809, physics.plasm-ph: 10456, physics.chem-ph: 10638, cond-mat.dis-nn: 11174,  nucl-ex: 11274, cond-mat: 11357, physics.soc-ph: 11630, physics.atom-ph: 11784, cond-mat.quant-gas: 13037, physics.ins-det: 13492, astro-ph.IM: 16781, physics.flu-dyn: 17354, hep-lat: 17449, astro-ph.EP: 20736, hep-ex: 22250, cond-mat.soft: 26552, physics.optics: 26949, cond-mat.supr-con: 30384, nucl-th: 32395, astro-ph.HE: 36665, astro-ph.CO: 38030, cond-mat.stat-mech: 39292, astro-ph.SR: 40994, astro-ph.GA: 43058, cond-mat.str-el: 45949, cond-mat.mtrl-sci: 56895, cond-mat.mes-hall: 60482, astro-ph: 94246, quant-ph: 102221, hep-th: 102314, \textbf{hep-ph: 128484}}

\textbf{Economics}
{econ.TH: 1377, econ.EM: 2112, \textbf{econ.GN: 2638}}

\textbf{Quantitative Biology}
{q-bio.SC: 651, q-bio.OT: 777, q-bio.CB: 911, q-bio.TO: 1077, q-bio.GN: 1667, q-bio.MN: 2128, q-bio.BM: 2629, q-bio.QM: 4439, q-bio.NC: 5529, \textbf{q-bio.PE: 6849}}

\textbf{Quantitative Finance}
{q-fin.EC: 384, q-fin.TR: 976, q-fin.PM: 1049, q-fin.CP: 1090, q-fin.RM: 1150, q-fin.PR: 1169, q-fin.MF: 1390, q-fin.GN: 1470, \textbf{q-fin.ST: 1828}}

\textbf{Statistics}
{stat.OT: 600, stat.CO: 3419, stat.AP: 8462, stat.ML: 15435, \textbf{stat.ME: 17378}}

\textbf{Computer Sciences}
{cs.GL: 106, cs.OS: 442, cs.MS: 980, cs.PF: 1040, cs.NA: 1083, cs.SC: 1170, cs.ET: 1857, cs.MM: 1939, cs.OH: 2002, cs.GR: 2179, cs.MA: 2280, cs.AR: 2531, cs.FL: 2693,cs.DL: 3165, cs.CE: 3271, cs.CG: 3943,cs.DM: 4408, cs.PL: 4479, cs.CC: 4786, cs.SY: 5130, cs.SD: 5397, cs.DB: 5487, cs.NE: 6011, cs.GT: 6678, cs.IR: 8590, cs.HC: 8696,cs.CY: 8947, cs.SI: 9236, cs.LO: 9690, cs.SE: 11333, cs.DC: 11981, cs.DS: 14338, cs.NI: 14662, cs.AI: 18871, cs.CR: 19266, cs.RO: 19594,cs.IT: 33285, cs.CL: 40190, cs.LG: 72867, cs.CV: 81633}

\textbf{Electrical Engineering and Systems Science}
{eess.AS: 5250, eess.SY: 11174, eess.IV: 12161, \textbf{eess.SP: 13722}}


\begin{figure}[!htbp]
\centering
\includegraphics[scale=0.25]{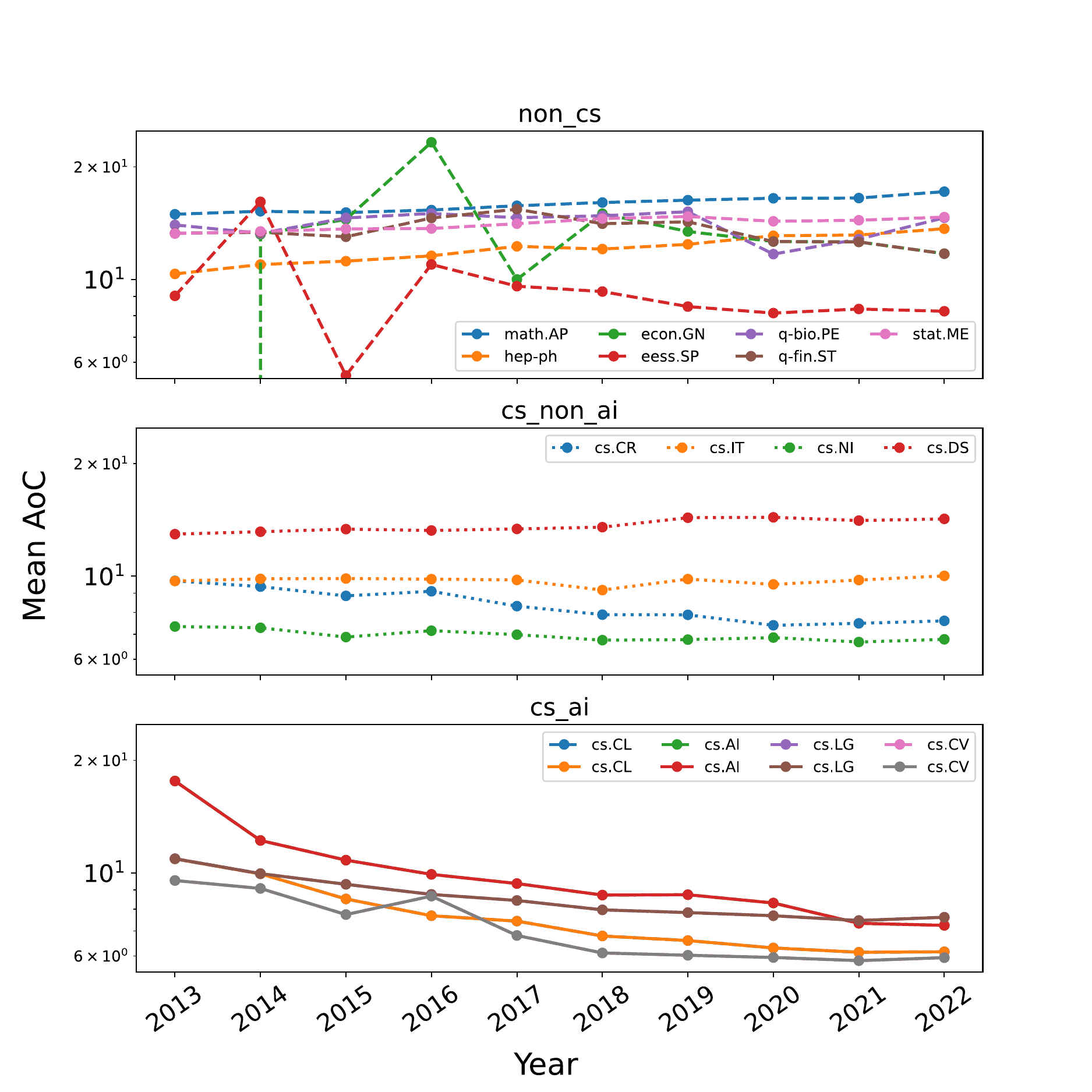}
\caption{Mean AoC of papers published from 2013 to 2022 by category. Log scale of y-axis.}
\label{fig:mean_aoc_log}
\centering
\end{figure}

\begin{figure}[!htbp]
\centering
\includegraphics[scale=0.25]{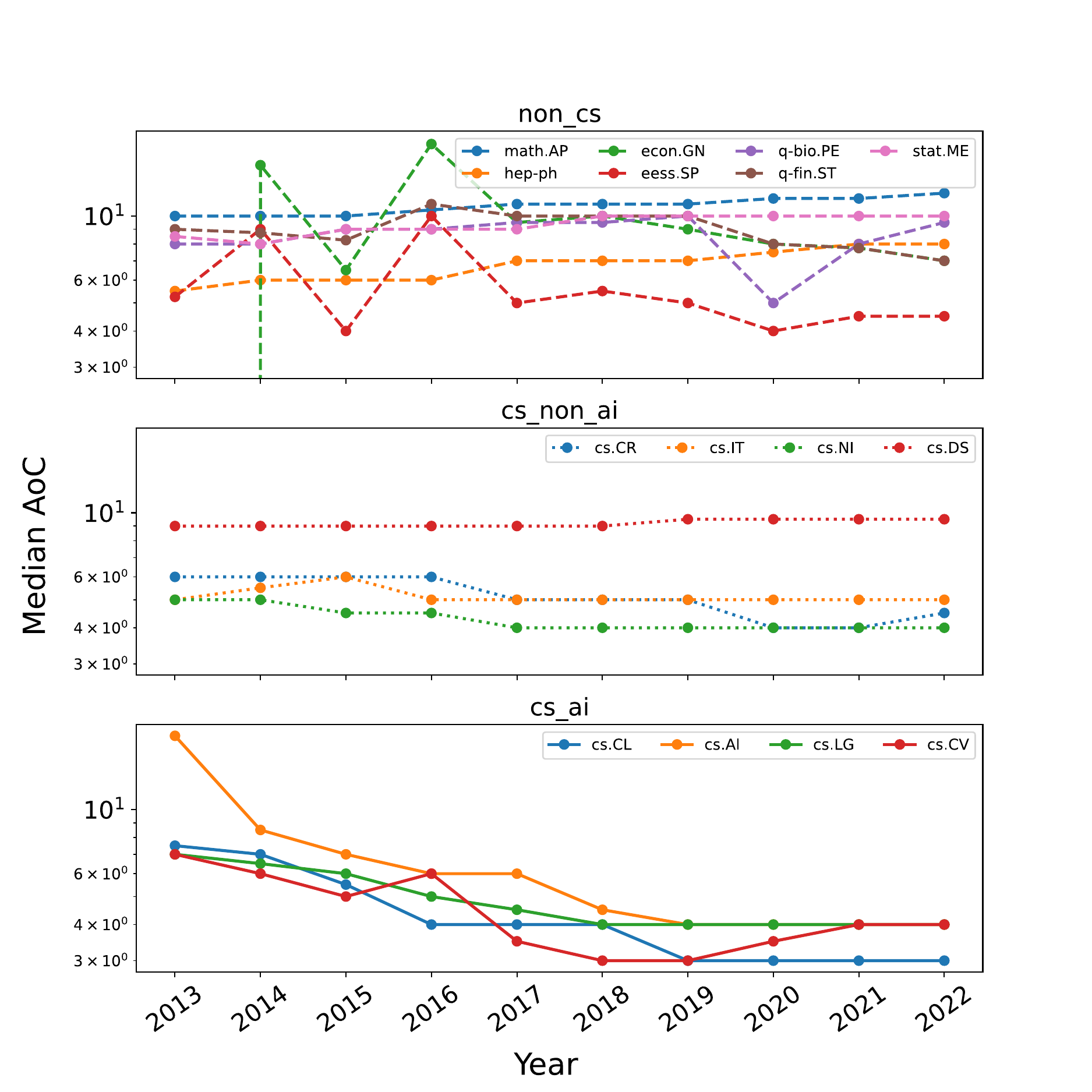}
\caption{Median AoC of papers published from 2013 to 2022 by category. Log scale of y-axis.}
\label{fig:meadian_aoc_log}
\centering
\end{figure}


\begin{table*}[htb]
\centering
\begin{tabular}{c|cccc|cccc}
     \toprule
     \textbf{year} & \textbf{cs.CR} & \textbf{cs.IT} & \textbf{cs.NI} & \textbf{cs.DS}  & \textbf{cs.AI} & \textbf{cs.CL} & \textbf{cs.CV} & \textbf{cs.LG} \\
     \midrule 
     2013    &  $460_{452}$      &  $1963_{1950}$     &   $778_{770}$    &  $704_{694}$   & $1383_{932}$      &   $219_{219}$    &  $662_{511}$     &  $696_{405}$  \\ 
     2014  &   $540_{529}$     &  $2013_{2007}$     &   $838_{830}$    &   $784_{775}$    &  $663_{487}$      &   $396_{355}$    &   $1096_{722}$    &   $739_{418}$ \\ 
     2015  &   $597_{529}$     & $2418_{2414}$      &  $766_{759}$     &   $873_{867}$    & $587_{355}$     & $587_{474}$      &   $1859_{1166}$    &   $1122_{537}$  \\ 
          2016  &   $715_{704}$     &  $2598_{2589}$     &  $994_{987}$     &  $978_{975}$ & $875_{554}$      &  $1306_{974}$     &   $3084_{1166}$    &   $1523_{717}$  \\ 
          2017  &   $1055_{1051}$     &  $2789_{2589}$     &   $1004_{997}$    &    $1053_{1050}$  &  $1209_{717}$      & $1922_{1425}$      &  $4914_{3012}$     &    $2340_{1122}$ \\ 
          2018  &  $1530_{1522}$      &  $2407_{2403}$     &  $1245_{1239}$     &    $1187_{1184}$   & $1442_{862}$      &  $2974_{2357}$     &  $7261_{4619}$     &  $4736_{2227}$  \\ 
          2019  &  $1937_{1910}$      &  $2250_{2246}$     &  $1295_{1289}$     &    $1302_{1296}$   &  $1210_{721}$      &  $4170_{3198}$     &   $8489_{6324}$    &   $8841_{4280}$   \\ 
          2020  &   $2393_{2382}$     &  $2355_{2344}$     &  $1337_{1323}$     &    $1441_{1436}$   &  $1916_{1126}$      &  $5582_{4109}$     &   $11,000_{8546}$    &  $11,097_{5214}$    \\ 
          2021  &  $2671_{2646}$      &  $2631_{2604}$     &  $1318_{1295}$     &    $1193_{1167}$   &  $2130_{1273}$      &  $6578_{4876}$     &   $12,695_{9849}$    &   $13,087_{6819}$ \\ 
          2022  &   $2843_{2806}$     &  $2541_{2511}$     &   $1313_{1297}$    &    $1237_{1216}$   & $ 2114_{1289}$      &  $7133_{5640}$     &   $14,625_{12476}$    &   $13,754_{7949}$  \\ 
     \bottomrule 
\end{tabular} 
\caption{Number of submissions in arXiv dataset on Kaggle. The numbers in subscripts are the actual numbers of publications in our dataset. Left: cs-non-ai categories. Right: cs-ai categories.}
\label{table:cs_submissions}
\end{table*}

\begin{table*}[htb]
\centering
\begin{tabular}{c|ccccccc}
     \toprule
     \textbf{year} & \textbf{q-bio.PE} & \textbf{q-fin.ST} & \textbf{stat.ME} & \textbf{hep-ph}  & \textbf{math.AP} & \textbf{econ.GN} & \textbf{eess.SP} \\
     \midrule 
     2013    &  $475_{471}$      &  $91_{91}$     &   $637_{629}$    &  $4642_{4193}$   & $2211_{4193}$ & $0_{0}$ & $2_{2}$\\ 
     2014  &   $402_{399}$     &  $98_{98}$     &   $765_{759}$    &   $4623_{4259}$    & $2467_{2467}$ & $2_{1}$ & $1_{1}$\\ 
          2015  &  $ 389_{388}$     & $91_{90}$      &  $919_{916}$     &    $4936_{4569}$  & $2835_{2827}$ & $2_{2}$ & $1_{1}$ \\ 
          2016  &   $402_{373}$     &  $99_{99}$     &  $1065_{1060}$     &    $4751_{4432}$  & $2820_{2811}$ & $2_{2}$ & $1_{1}$ \\ 
          2017  &   $386_{384}$     &  $85_{85}$     &   $1319_{1314}$    &    $4516_{4287}$  & $3202_{3194}$ & $4_{4}$ & $331_{331}$ \\ 
          2018  &  $386_{382}$      &  $112_{112}$     &  $1313_{1306}$     &    $4571_{4447}$ & $3476_{3461}$ & $118_{117}$ & $1662_{1652}$  \\ 
          2019  &  $390_{386}$      &  $137_{137}$     &  $1356_{1351}$     &    $4574_{4445}$  & $3670_{3667}$ & $241_{240}$ & $2242_{2231}$ \\ 
          2020  &   $999_{974}$     &  $191_{190}$     &  $1962_{1955}$     &   $4598_{4454}$ & $4006_{3990}$ & $450_{190}$ & $29964_{2951}$  \\ 
          2021  &  $542_{534}$      &  $179_{176}$     &  $2160_{2134}$     &   $4979_{4724}$  & $3882_{3798}$ & $692_{176}$ & $2964_{2951}$ \\ 
          2022  &   $426_{417}$     &  $155_{153}$     &   $2279_{2243}$   &    $5174_{3506}$  & $3952_{3815}$ & $601_{153}$ & $2339_{2314}$ \\ 
     \bottomrule
\end{tabular}
\caption{Number of submissions in arXiv dataset on Kaggle. The numbers in subscripts are the actual numbers of publications in our dataset. non-cs categories.}
\label{table:noncs_submissions}
\end{table*} 


\begin{table*}[htb]
\centering
\begin{tabular}{c|cccc|cccc}
     \toprule
     \textbf{year} & \textbf{cs.CR} & \textbf{cs.IT} & \textbf{cs.NI} & \textbf{cs.DS}  & \textbf{cs.AI} & \textbf{cs.CL} & \textbf{cs.CV} & \textbf{cs.LG} \\
     \midrule 
     2013    &  6 &  5 &  5  &  9 &  18 &  7.5 &  7 &  7  \\ 
     2014  &   6 &  5.5 &  5 &  9 &  8.5 &  7 &  6 &  6.5 \\ 
          2015  &  6 &  6 &  4.5 &  9 &  7 &  5.5 &  5 &  6 \\ 
          2016  &  6 &  5 &  4.5 &  9 &  6 &  4 &  6 &  6 \\ 
          2017  & 5 &  5 &  4 &  9 &  6 &  4 &  3.5 &  5\\ 
          2018  & 5 &  5 &  4 &  9 &  4.5 &  4 &  3 &  4.5 \\ 
          2019  & 5 &  5 &  4 &  	9.5 &  4 &  3 &  3 &  4 \\ 
          2020  & 4 &  5 &  4 &  9.5 &  4 &  3 &  3.5 &  4 \\ 
          2021  & 4 &  5 &  4 &  9.5 &  4 &  3 &  4 &  4\\ 
          2022  & 4.5 &  5 &  4 &  9.5 &  4 &  3 &  4 &  4 \\ 
     \bottomrule 
\end{tabular} 
\caption{Median AoC. Left: cs-non-ai categories. Right: cs-ai categories.}
\label{table:csmedian}
\end{table*}

\begin{table*}[htb]
\centering
\begin{tabular}{c|ccccccc}
     \toprule
     \textbf{year} & \textbf{q-bio.PE} & \textbf{q-fin.ST} & \textbf{stat.ME} & \textbf{hep-ph}  & \textbf{math.AP} & \textbf{econ.GN} & \textbf{eess.SP} \\
     \midrule 
     2013    & 8 & 9 & 8.5 & 5.5 & 10 & 0 & 5.25\\ 
     2014  &   8 & 8.75 & 8 & 6 & 10 & 15 & 9\\ 
          2015  & 9 & 8.25 & 9 & 6 & 10 & 6.5 & 4\\ 
          2016  & 9 & 11 & 9 & 6 & 10.5 & 17.75 & 10\\ 
          2017  &  9.5 & 10 & 9 & 7 & 11 & 9.5 & 5\\ 
          2018  & 9.5 & 10 & 10 & 7 & 11 & 10 & 5.5 \\ 
          2019  &  10 & 10 & 10 & 7 & 11 & 9 & 5 \\ 
          2020  &  5 & 8 & 10 & 7.5 & 11.5 & 8 & 5 \\ 
          2021  &  8 & 7.75 & 10 & 8 & 11.5 & 7.75 & 4.5 \\ 
          2022  &  9.5 & 7 & 10 & 8 & 12 & 7 &  4.5\\ 
     \bottomrule
\end{tabular}
\caption{Median AoC of non-cs categories.}
\label{table:noncsmedian}
\end{table*} 

\begin{table*}[htb]
\centering
\begin{tabular}{c|cccc|cccc}
     \toprule
     \textbf{year} & \textbf{cs.CR} & \textbf{cs.IT} & \textbf{cs.NI} & \textbf{cs.DS}  & \textbf{cs.AI} & \textbf{cs.CL} & \textbf{cs.CV} & \textbf{cs.LG} \\
     \midrule 
     2013    & 9.8\% &	8.09\% &	6.63\% &	14.59\% &	30.15\% &	14.04\% &	10.34\% &	13.36\%  \\ 
     2014  &   9.97\% &	8.52\% &	6.59\% &	14.78\% &	17.07\% &	12.87\% &	9.81\% &	10.93\%     \\ 
          2015  &   8.78\% &	8.5\% &	6.16\% &	14.56\% &	13.43\% &	9.48\% &	9.81\% &	9.72\%     \\ 
          2016  &   9.67\% &	8.63\% &	6.61\% &	14.5\% &	11.2\% &	9.48\% &	8.46\% &	8.45\%    \\ 
          2017  &  7.70\%    &  8.28\%     &   6.26\%    &    14.87\%  &  10.09\%     & 8.03\%     &  5.39\% & 6.77\%    \\ 
          2018  & 7.05\% &	7.5\% &	5.59\% &	14.91\%	 & 8.57\% &	5.11\%    &	4.07\%    &	6.91\%   \\ 
          2019  &  7.05\% &	8.71\% &	5.38\% &	15.52\% &	8.46\% &	4.78\% &	3.57\% &	6.45\%    \\ 
          2020  &   6.33\% &	7.76\% &	5.36\% &	15.97\% &	7.23\% &	4.31\% &	3.22\% &	5.9\%    \\ 
          2021  &  5.9\% &	8.26\% &	5.01\% &	16\% &	5.19\% &	3.67\% &	2.63\% &	5.12\%   \\ 
          2022  &  5.83\% &	8.61\% &	5.01\% &	3.67\% &	4.59\% &	3.36\% &	2.48\% &	4.86\%    \\ 
     \bottomrule 
\end{tabular} 
\caption{Percentage of old papers. Left: cs-non-ai categories. Right: cs-ai categories.}
\label{table:csold}
\end{table*}

\begin{table*}[htb]
\centering
\begin{tabular}{c|ccccccc}
     \toprule
     \textbf{year} & \textbf{q-bio.PE} & \textbf{q-fin.ST} & \textbf{stat.ME} & \textbf{hep-ph}  & \textbf{math.AP} & \textbf{econ.GN} & \textbf{eess.SP} \\
     \midrule 
     2013    & 12.79\% &	15.32\% &	14.16\% &	9.18\% &	17.21\% &	0\% &	4.78\%\\ 
     2014  &   11.81\% &	15.64\% &	13.8\% &	9.6\% & 17.32\% &	40\% &	11.95\%\\ 
          2015  &   13.83\% &	14.32\% &	15.07\% &	9.55\% &	16.9\% &	16.74\% &	8\% \\ 
          2016  &   15.18\% &	18.64\% &	14.83\% &	9.64\% &	17.59\% &	27.57\% &	19.35\% \\ 
          2017  &   14.77\% &	15.35\% &	14.9\% &	9.95\% &	17.86\% &	12.05\% &	8.77\%   \\ 
          2018  & 15.11\% &	16.88\% &	16.07\% &	10.64\% &	18.18\% &	17.37\% &	9.2\% \\ 
          2019  & 15.29\% &	17.62\% &	16.15\% &	11.07\% &	18.42\% &	15.21\% &	7.43\% \\ 
          2020  & 10.92\% &	14.45\% &	15.86\% &	11.3\% &	19.04\% &	14.45\% &	6.95\% \\ 
          2021  &  12.84\% &	13.72\% &	16.29\% &	11.23\% &	19.22\% &	13.72\% &	7.04\% \\ 
          2022  &  15.39\% &	13.37\% &	16.78\% &	11.27\% &	20.11\% &	13.37\% &	7.14\% \\ 
     \bottomrule
\end{tabular}
\caption{Percentage of old papers of non-cs categories.}
\label{table:noncsold}
\end{table*} 

\begin{table*}[htb]
\centering
\begin{tabular}{c|cccc|cccc}
     \toprule
     \textbf{year} & \textbf{cs.CR} & \textbf{cs.IT} & \textbf{cs.NI} & \textbf{cs.DS}  & \textbf{cs.AI} & \textbf{cs.CL} & \textbf{cs.CV} & \textbf{cs.LG} \\
     \midrule 
     2013    &  8.77 & 9.14 & 7.25 & 11.06 & 16.11 & 11.25 & 8.56 & 9.9 \\ 
     2014  & 8.41 & 9.14 & 6.77 & 11.13 & 11.67 & 9.12 & 7.87 & 8.9 \\ 
          2015  & 7.9 & 9.22 & 6.33 & 11.22 & 10.23 & 7.96 & 6.59 & 8.35 \\ 
          2016  &  8.26 & 9.4 & 7 & 11.75 & 8.77 & 6.7 & 7.55 & 7.39 \\ 
          2017  & 7.37 & 9.18 & 6.9 & 11.75 & 8.2 & 6.32 & 5.69 & 7.27 \\ 
          2018  & 7.07 & 8.42 & 6.29 & 12.04 & 7.49 & 5.9 & 5.01 & 6.97\\ 
          2019  & 7.43 & 9.13 & 6.9 & 12.54 & 7.69 & 5.55 & 5.01 & 7.01 \\ 
          2020  & 6.9 & 9.13 & 6.85 & 12.99 & 7.29 & 5.27	 & 5.05 & 6.86 \\ 
          2021  & 7 & 8.96 & 6.05 & 12.65 & 	6.5 & 5.14 & 5.1 & 6.63 \\ 
          2022  & 6.9	 & 9.36 & 6.67 & 13.3 & 6.4 & 5.29 & 5.1 & 6.89\\ 
     \bottomrule 
\end{tabular} 
\caption{Mean AoC of influential citations. Left: cs-non-ai categories. Right: cs-ai categories.}
\label{table:csinfluential}
\end{table*}

\begin{table*}[htb]
\centering
\begin{tabular}{c|ccccccc}
     \toprule
     \textbf{year} & \textbf{q-bio.PE} & \textbf{q-fin.ST} & \textbf{stat.ME} & \textbf{hep-ph}  & \textbf{math.AP} & \textbf{econ.GN} & \textbf{eess.SP} \\
     \midrule 
     2013    & 12.48 & 12.03 & 12.49 & 10.18 & 	13.86 & 0 & 8.25\\ 
     2014  &   11.52 & 11.65 & 12.35 & 9.71 & 14.56 & 0 & 17.33\\ 
          2015  & 13.84 & 13.01 & 12.63 & 10.32 & 14.27 & 11.83 & 3.33\\ 
          2016  & 14.5 & 14.84 & 13.01 & 10.32 & 	15.09 & 22.29 & 12.9 \\ 
          2017  & 14.14 & 14.76 & 13.52 & 11.61 & 15.19 & 8 & 8.61\\ 
          2018  & 13.95 & 15.49 & 13.52 & 11.12 & 15.77 & 14.88 & 8.48 \\ 
          2019  & 14.36 & 13.89 & 14.12 & 11.8 & 16.23 & 12.37 & 7.81\\ 
          2020  & 10.69 & 12.68 & 13.76 & 12.17 & 16 & 12.68 & 7.61 \\ 
          2021  & 12.33 & 13.3 & 13.61 & 11.97 & 15.69 & 13.3 & 8.05\\ 
          2022  & 13.37 & 11.67 & 13.95 & 11.73 & 16.8 & 11.67 & 7.78\\ 
     \bottomrule
\end{tabular}
\caption{Mean AoC of influential citations of non-cs categories.}
\label{table:noncsinfluential}
\end{table*}
\end{appendices}
\end{document}